\documentclass[pdflatex,sn-mathphys,Numbered,iicol]{sn-jnl}

\usepackage{graphicx}%
\usepackage{multirow}%
\usepackage{amsmath,amssymb,amsfonts}%
\usepackage{amsthm}%
\usepackage{mathrsfs}%
\usepackage[title]{appendix}%
\usepackage{xcolor}%
\usepackage{textcomp}%
\usepackage{manyfoot}%
\usepackage{booktabs}%
\usepackage{algorithm}%
\usepackage{algorithmicx}%
\usepackage{algpseudocode}%
\usepackage{listings}%

\usepackage[normalem]{ulem}

\begin{document}
\title{Instrumental uncertainties in radiative corrections for the MUSE experiment}


\author[1]{\fnm{L.} \sur{Li}}
\author*[1]{\fnm{S.} \sur{Strauch}}\email{strauch@sc.edu}
\author[2,3]{\fnm{J. C.} \sur{Bernauer}}
\author[4]{\fnm{W. J.} \sur{Briscoe}}
\author[5]{\fnm{A.} \sur{Christopher Ndukwe}}
\author[2]{\fnm{E.} \sur{Cline}}
\author[6]{\fnm{D.} \sur{Cohen}}
\author[7]{\fnm{K.} \sur{Deiters}}
\author[4]{\fnm{E. J.} \sur{Downie}}
\author[8,5]{\fnm{I. P.} \sur{Fernando}}
\author[1]{\fnm{A.} \sur{Flannery}}
\author[9]{\fnm{R.} \sur{Gilman}}
\author[1]{\fnm{Y.} \sur{Ilieva}}
\author[5]{\fnm{M.} \sur{Kohl}}
\author[4,10]{\fnm{I.} \sur{Lavrukhin}}
\author[9]{\fnm{W.} \sur{Lin}}
\author[10]{\fnm{W.} \sur{Lorenzon}}
\author[10]{\fnm{S.} \sur{Lunkenheimer}}
\author[13]{\fnm{P.} \sur{Mohanmurthy}}
\author[5]{\fnm{J.} \sur{Nazeer}}
\author[1]{\fnm{M.} \sur{Nicol}}
\author[5]{\fnm{T.} \sur{Patel}}
\author[6]{\fnm{A.} \sur{Prosnyakov}}
\author[10]{\fnm{H.} \sur{Reid}}
\author[11]{\fnm{P. E.} \sur{Reimer}}
\author[6]{\fnm{G.} \sur{Ron}}
\author[7]{\fnm{T.} \sur{Rostomyan}}
\author[6]{\fnm{O. M.} \sur{Ruimi}}
\author[12]{\fnm{N.} \sur{Sparveris}}
\author[6]{\fnm{D.} \sur{Yaari}}

\affil[1]{\orgdiv{Department of Physics and Astronomy}, \orgname{University of South Carolina}, \orgaddress{\city{Columbia}, \state{South Carolina} \postcode{29208}, \country{USA}}}
\affil[2]{\orgdiv{Department of Physics and Astronomy}, \orgname{Stony Brook University}, \orgaddress{\city{Stony Brook}, \state{New York} \postcode{11794},  \country{USA}}}
\affil[3]{\orgdiv{RIKEN BNL Research Center}, \orgname{Brookhaven National Laboratory}, \orgaddress{\city{Upton}, \state{New York} \postcode{11973}, \country{USA}}}
\affil[4]{\orgdiv{Department of Physics}, \orgname{The George Washington University}, \orgaddress{\city{Washington}, \state{D.C.} \postcode{20052}, \country{USA}}}
\affil[5]{\orgdiv{Department of Physics}, \orgname{Hampton University}, \orgaddress{\city{Hampton}, \state{Virginia} \postcode{23668}, \country{USA}}}
\affil[6]{\orgdiv{Racah Institute of Physics}, \orgname{The Hebrew University of Jerusalem}, \orgaddress{\city{Jerusalem} \postcode{91904}, \country{Israel}}}
\affil[7]{\orgname{Paul Scherrer Institute}, \orgaddress{\postcode{CH-5232} \city{Villigen}, \country{Switzerland}}}
\affil[8]{\orgdiv{Department of Physics}, \orgname{University of Virginia}, \orgaddress{\city{Charlottesville}, \state{Virginia} \postcode{22904}, \country{USA}}}
\affil[9]{\orgdiv{Department of Physics and Astronomy}, \orgname{Rutgers, The State University of New Jersey}, \orgaddress{\city{Piscataway}, \state{New Jersey} \postcode{08854}, \country{USA}}}
\affil[10]{\orgdiv{Randall Laboratory of Physics}, \orgname{University of Michigan}, \orgaddress{\city{Ann Arbor}, \state{Michigan} \postcode{48109}, \country{USA}}}
\affil[11]{\orgdiv{Physics Division}, \orgname{Argonne National Laboratory}, \orgaddress{\city{Lemont}, \state{Illinois} \postcode{60439}, \country{USA}}}
\affil[12]{\orgdiv{Department of Physics}, \orgname{Temple University}, \orgaddress{\city{Philadelphia}, \state{Pennsylvania} \postcode{19122}, \country{USA}}}
\affil[13]{\orgdiv{Laboratory for Nuclear Science}, \orgname{Massachusetts Institute of Technology}, \orgaddress{\city{Cambridge}, \state{Massachusetts} \postcode{02139}, \country{USA}}}


\abstract{The MUSE experiment at the Paul Scherrer Institute is measuring elastic lepton-proton scattering cross sections in a four-momentum transfer range from \textit{Q}\textsuperscript{2} of approximately 0.002 to 0.08 GeV\textsuperscript{2} using positively and negatively charged electrons and muons. The extraction of the Born cross sections from the experimental data requires radiative corrections. Estimates of the instrumental uncertainties in those corrections have been made using the ESEPP event generator.  The results depend in particular on the minimum lepton momentum that contributes to the experimental cross section and the fraction of events with hard initial-state radiation that is detected in the MUSE calorimeter and is excluded from the data. These results show that the angular-dependent instrumental uncertainties in radiative corrections to the electron cross section are less than 0.4~\% and are negligible for the muon cross section.}

\keywords{Electromagnetic interactions, Photon and charged-lepton interactions with hadrons, Elastic and Compton scattering}

\maketitle

\section{Introduction}

The MUon Scattering Experiment (MUSE) at the Paul Scherrer Institute (PSI)~\cite{Gilman:2017hdr,Cline:2021ehf} has been developed to measure elastic electron-proton and muon-proton scattering cross sections of positively and negatively charged leptons. Measurements are done at beam momenta $p_0$ of 115 MeV/$c$, 161 MeV/$c$, and 210 MeV/$c$ and over a wide range of scattering angles $\theta$ between 20$^\circ$ and 100$^\circ$.  MUSE covers four-momentum-transfers from $Q^2$ of approximately 0.002 to 0.08 GeV$^2$. Each of the four data sets , $e^\pm p$ and $\mu^\pm p$, allows the extraction of the proton charge radius. In combination, the data test possible differences between the electron and muon interactions and two-photon exchange effects. 

For each beam momentum, electron and muon scattering cross sections, $(d\sigma/d\Omega_\ell)_\text{exp}$, will be obtained as a function of the lepton scattering angle. The extraction of the electromagnetic form factors from the measured cross sections will use an iterative approach to match simulated and measured count rates. Radiative effects can be quantified by a correction factor $(1+\delta)$, which is defined in the relation between the measured experimental cross section and the Born cross section \cite{Gramolin:2014pva}%
\begin{equation}
\left(\frac{d\sigma}{d\Omega_\ell}\right)_\text{exp}=\left(\frac{d\sigma}{d\Omega_\ell}\right)_{\text{Born}}(1+\delta).
\end{equation}
The radiative correction accounts for higher-order processes, such as the dominant first-order bremsstrahlung process $\ell p\to \ell'p\gamma$, the vacuum polarization correction, the vertex corrections, and the two-photon-exchange corrections. It is estimated from model calculations of the Born and bremsstrahlung, $d\sigma_\text{brems}$, cross sections in the kinematic settings of the experiment. Monte-Carlo simulations are used to carry out the integration of the bremsstrahlung cross section
\begin{equation}
\left(\frac{d\sigma}{d\Omega_\ell}\right)_\text{exp} = \int_{p'_\ell} \int_{\Omega_\gamma} \frac{d\sigma_\text{brems}}{d\Omega_\ell d\Omega_\gamma dp'_\ell}d\Omega_\gamma dp_\ell',
\label{eq:int}
\end{equation}
where $p'_\ell$ is the momentum of the scattered lepton, and $\Omega_\ell$ and $\Omega_\gamma$ are the solid angles of the incident lepton and final-state photon, respectively. The limits in the integration reflect the experimental conditions. The kinematics, event selection, lepton, and photon angular and momentum acceptances, detector resolutions, and functional behavior of the form factors will thus affect the size of the radiative corrections.    

Dedicated calculations of radiative corrections for MUSE have been performed  by Afanasev and collaborators~\cite{Afanasev:2021nhy, Koshchii:2017dzr, Afanasev:2020ejr}, by Myhrer and collaborators in the framework of the Heavy Baryon Chiral Perturbation Theory~\cite{Talukdar:2018hia, Talukdar:2019dko}, and within the McMule framework using the higher-order QED calculations ~\cite{Ulrich:2020frs, Banerjee:2020rww}. Second-order radiative corrections have also been calculated for MUSE in Ref.~\cite{Bucoveanu:2018soy}. Two-photon exchange effects in MUSE kinematical region were studied by Tomalak and Vanderhaeghen~\cite{Tomalak:2014dja,Tomalak:2015hva,Tomalak:2018ere,Tomalak:2018jak} in the dispersion relation approach, by Weiss and collaborators~\cite{Gil-Dominguez:2023hku} in the framework of Dispersively Improved Chiral Effective Field Theory, and by Paz and collaborators~\cite{Dye:2018rgg} using QED-NRQED effective field theory. However, the simulations to perform the integral in Eq.~(\ref{eq:int}) require a suitable event generator, which, in the case of MUSE, must fulfill several requirements. First, the event generator needs to include the emission of a hard radiated photon in the initial and final state, which is beyond the soft-photon approximation. MUSE includes a large fraction of the radiative tail in the momentum acceptance. The hard photon needs to be propagated in the full Monte Carlo simulation. Second, the event generator needs to include the mass of the leptons to avoid the $Q^2 \gg m^2$ approximation in the calculation. The Elastic Scattering of Electrons and Positrons on Protons (ESEPP) event generator \cite{Gramolin:2014pva} takes into account the first-order radiative corrections of elastic scattering of charged leptons ($e$$^{\pm}$ and $\mu$$^{\pm}$) on protons and fulfills these requirements. We have used ESEPP with the Kelly parametrization of the proton form factors~\cite{Kelly:2004hm} for the present studies.  Other generators that are viable for MUSE include those discussed in  Refs.~\cite{Akushevich:2011zy, A1:2013fsc}, and were found to give results in good agreement with those from ESEPP when run under similar conditions.

This paper presents a study of how the instrumental characteristics of the MUSE experiment impact the determination of radiative corrections. The details of the Monte Carlo simulation and the properties of detectors related to the radiative corrections will be presented. Results of the radiative corrections and their uncertainties from preliminary instrumental input to the model calculations will be discussed.

\section{Experiment}

\subsection{Setup}

\begin{figure}[ht]
  \centering
  \includegraphics[width=\columnwidth]{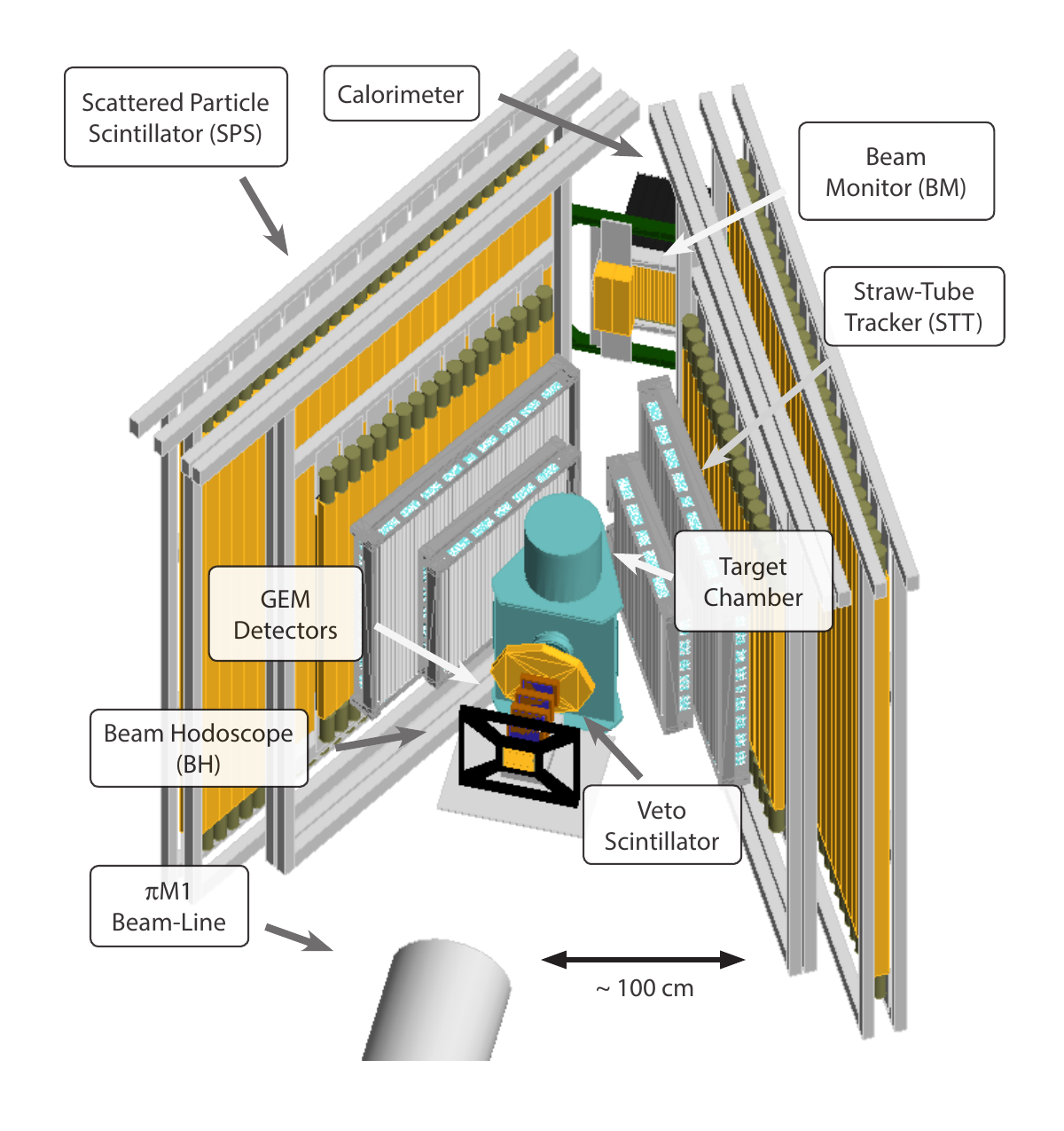}
\caption{Sketch of the MUSE experimental setup at the $\pi$M1 beamline at PSI as implemented in the MUSE Geant4 simulation.}
\label{fig:muse_timing_2020920}
\end{figure}

Figure~\ref{fig:muse_timing_2020920} shows a schematic layout of the experimental setup of MUSE in the $\pi$M1 secondary beamline \cite{Cline:2021vlw} at PSI. 
The particle beam contains a mix of electrons, muons, and pions with a beam flux up to 3.5~MHz in the experiment. These particles go through the Beam Hodoscope (BH) detector for timing and particle identification \cite{Rostomyan:2020tnt}. The BH also measures the beam fluxes.  Three Gas Electron Multiplier (GEM) chambers~\cite{Liyanage:2018wup} determine the incident particle track.  A veto detector suppresses triggers from off-axis particles. The target chamber contains a ladder with liquid-hydrogen, empty-cell, and solid targets, in addition to a non-target position \cite{Roy:2019add}. Scattered particles are detected by two symmetric spectrometers, each with two Straw-Tube Tracker (STT) chambers and two planes of fast Scattered Particle Scintillators (SPS). The combined information from the GEM and STT trackers allow for the reconstruction of the vertex and scattering angle for the $ep$ or $\mu p$ reaction. The scattering-angular resolution for one event is dominated by multiple scattering and is smaller than 20~mrad. Systematic uncertainties of the scattering angle reconstruction are expected to be below $\sigma_\theta=1$~mrad. The two SPS planes (front-wall and rear-wall) provide the scattered-particle event trigger which requires a hit in each scintillator wall. The unscattered particles pass the Beam Monitor (BM). The time-of-flight (TOF) from the BH to the SPS determines the reaction type (muon scattering vs. muon decay in flight). The TOF from the BH to the BM determines $\mu$ and $\pi$ beam momenta. Channel beam momenta have been verified with those measurements to within $\sigma_{p_0}/p_0 = 0.2$ \%~\cite{Cline:2021vlw}. The BM can also be used to suppress background from M\o ller and Bhabha scattering. The most downstream beamline detector is the calorimeter. It consists of 64 lead-glass blocks in a square arrangement 139~cm downstream of the target.  The blocks have a size of $4\ \text{cm}\ \times\ 4\ \text{cm}\ \times 30\ \text{cm}$.  The calorimeter is used to detect photons in the beam direction. 

Particularly important for the determination of the radiative corrections are the lepton-detection threshold in the SPS detectors and the hard photon detection in the calorimeter.

\subsection{Event selection}

\begin{figure}[htbp]
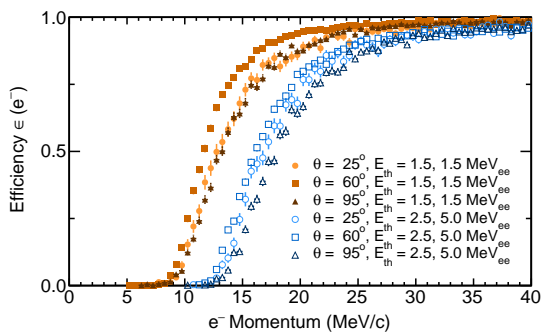

  \centering
  \includegraphics[width=\columnwidth]{{{fig/em_40MeV_full_full4pi_momdist_1.5MeV_1.5MeV_pmin_Combo_n_one}}}
\caption{Geant4 simulation of the electron detection threshold of the MUSE SPS detector.  The efficiency of particle detection is shown as a function of momentum for various scattering angles and low (full symbols) and high (open symbols) light-output thresholds.  The light-output thresholds $E_\text{th}$ are given for the front and rear-walls separately.}
\label{fig:efficiency}
\end{figure}

The magnitude of the momentum of the final-state lepton is not directly measured in MUSE. Instead, all scattered leptons, from the threshold, $p'_{\ell, \text{min}}$, to the elastic endpoint are included in the experimental yield. The threshold momentum depends on the scintillator light output and the discrimination thresholds set in the front- and rear-wall detectors of the SPS.\footnote{The electron light-output function of the SPS plastic scintillators is very nearly linear, and has been calibrated with photon sources following the method of Ref.~\cite{DIETZE1982549}. The light output is reported in units of MeV electron equivalent (MeV$_{ee}$).}  A Geant4 simulation of the electron detection efficiency, $\epsilon(p'_e)$, is shown in Fig.~\ref{fig:efficiency}. 
The full symbols are for light output thresholds of 1.5~MeV$_\text{ee}$ for each of the front and rear walls.  The open symbols are for higher threshold values of 2.5~MeV$_\text{ee}$ and 5.0~MeV$_\text{ee}$ in the two walls, respectively.  Particles with a scattering angle of about 60$^\circ$ traverse the walls approximately perpendicularly.  Smaller or larger scattering angles result in higher momentum thresholds, as shown in Fig.~\ref{fig:efficiency}, due to the larger path length in the scintillation material and related larger energy losses in the front wall. At very low momenta the lepton stops in the front wall, or deposits insufficient energy for the light output to be above the discriminator threshold.

The effective minimum momentum threshold, $p'_\text{min}$, was determined to be the momentum value for which the integral of the bremsstrahlung cross section over the detection range from $p'_\text{min}$ to the endpoint, $p'_\text{max}$, matches the efficiency weighted integral over all momenta

\begin{multline}
        \int_{p'_\text{min}}^{p'_\text{max}} \frac{d\sigma_\text{brems}}{d\Omega_\ell d\Omega_\gamma dp'_\ell}d\Omega_\gamma dp_\ell' = \\
        \int_0^{p'_\text{max}} \epsilon(p'_\ell)\frac{d\sigma_\text{brems}}{d\Omega_\ell d\Omega_\gamma dp'_\ell}d\Omega_\gamma dp_\ell'.
        \label{eq:pmin}
\end{multline}

We have evaluated Eq.~(\ref{eq:pmin}) using $\epsilon(p'_\ell)$ from Fig.~\ref{fig:efficiency} and the ESEPP event generator for $\sigma_\text{brems}$. The momentum threshold is a function of the discriminator thresholds in the SPS, the lepton scattering angle, and details of the event selection, e.g., cuts on the bremsstrahlung photons.  For electrons and detection thresholds of 2.0~MeV$_\text{ee}$ in both walls, it is about $p'_\text{min} = 14$~MeV/$c$, and for muons about $p'_\text{min} = 84$~MeV/$c$.  We estimate the uncertainty of the threshold values to be $\sigma_{p'_\text{min}}=2$~MeV/$c$.

\begin{figure}[htbp]
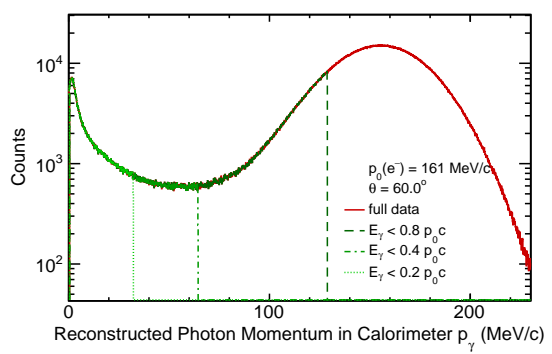

  \centering
  \includegraphics[width=\columnwidth]{{{fig/figures_e-_161.0MeV_060.0deg_Fig.6_photon_p_cal}}}
\caption{Simulation of the reconstructed photon momentum in the MUSE calorimeter (red histogram) for scattering of 161 MeV/$c$ electrons at 60$^\circ$.
}
\label{fig:cal}
\end{figure}

Photon production is often ignored in the experimental yield in electron-scattering experiments, and the bremsstrahlung cross section is integrated over the full solid angle of the real photons in the evaluation of radiative corrections. However, as shown below in Sec.~\ref{sec:results}, uncertainties in the radiative corrections in MUSE can be reduced when hard photons in the forward direction are excluded in the cross section. As an example of the anticipated MUSE calorimeter response, Fig.~\ref{fig:cal} 
shows the reconstructed photon momentum distribution for the scattering of 161 MeV/$c$ electrons at 60$^\circ$ as a red histogram. The simulation is based on the ESEPP event generator. It includes the geometrical acceptance of the $8 \times 8$ calorimeter blocks, and the energy-dependent detector resolution. The parameterization of the resolution was tuned to experimental data with electron-beam momenta in the range from 20~MeV/$c$ to 230~MeV/$c$.  When the calorimeter is used to suppress initial-state radiation, only events with a reconstructed photon energy less than a specified energy cut are used in the subsequent analysis.  For example, the green histograms in Fig.~\ref{fig:cal} display the accepted data with photon energies less than 80~\%, 40~\%, and 20~\% of the beam energy.  Preliminary commissioning data show a relative energy resolution of the calorimeter of 20~\%, at the energy corresponding to about half the beam momentum of 210~MeV/$c$, up to 25~\% for 115~MeV/$c$. We assume the uncertainty of the {\em mean} of the reconstructed photon-energy to be better than $\sigma_{E_\gamma}=5$~MeV.

\section{Simulations of the lepton-scattering cross section}
\label{sec:tail}

\begin{figure}[htbp]
  \centering
  \includegraphics[width=\columnwidth]{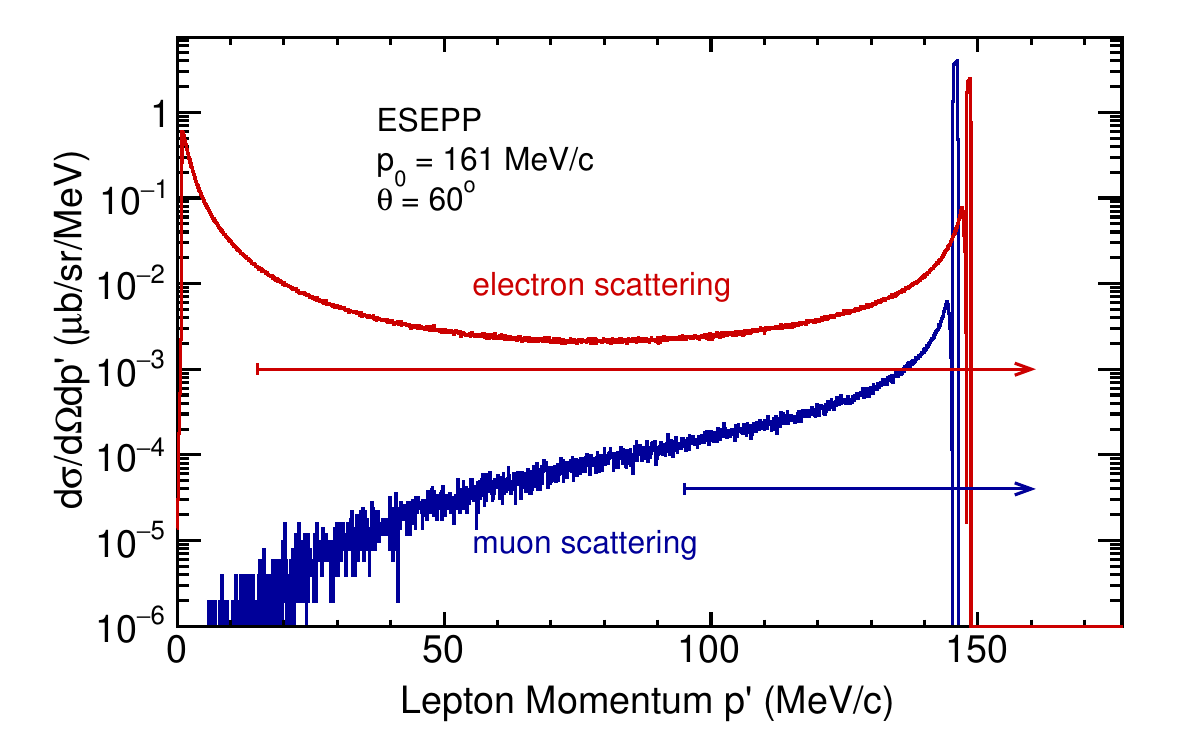}
\caption{ESEPP calculations of the electron (red) and muon (blue) scattering cross sections for an incident beam momentum of 161 MeV/$c$ and a scattering angle of 60$^\circ$.  The cross sections are integrated over all final-state photon directions.  The horizontal lines indicate the lepton-momentum acceptance of MUSE.}
\label{fig:tail}
\end{figure}

Figure~\ref{fig:tail} shows histograms from ESEPP simulations of the electron (red) and muon (blue) scattering cross sections at a kinematic setting with an incident beam momentum of $p_0=161$ MeV/$c$ and scattering angle $\theta$ of $60^\circ$. This setting is at the center of the momentum and angular range of the experiment.
In the experiment, all scattered particles above the momentum detection threshold $p'_\text{min}$ contribute to the experimental yield.  The momentum ranges are indicated by the horizontal lines. Each histogram shows the elastic peak at high momentum and its radiative tail. The muon elastic peak is slightly offset from the electron peak due to the larger proton recoil energy in $\mu p$ scattering. The apparent gap between the elastic peak and the tail is an artifact of the ESEPP calculation.  The event generator includes explicitly hard bremsstrahlung photons in the tail region with  energies as low as the cut-off energy $E_\gamma^\text{cut}$. The contribution to the cross section from photons with $E_\gamma < E_\gamma^\text{cut}$ is integrated over all photon directions and energies and is included in the elastic peak \cite{Gramolin:2014pva}.  Due to their small mass, the cross section in the radiative tail is much larger for electrons than muons.  In this kinematic setting, the cross section for $ep$ at the $e$ detection threshold is about 200 times larger than for $\mu p$ at the $\mu$ threshold.  The $\ell p\to \ell'p\gamma$ cross section vanishes for both lepton species as the scattered-lepton momentum reaches zero.  The local maximum in the electron radiative tail at low final-state electron momenta is caused by the increase of the $ep$ scattering cross section after initial-state photon emission from the electron in the beam direction and the subsequent reduction of the electron momentum and momentum transfer in the scattering process~\cite{Talukdar:2018hia}.  

\section{Radiative corrections}
\label{sec:results}

We have used the ESEPP event generator in a simplified simulation of MUSE and studied the radiative corrections $\delta$ in a variety of experimental conditions. The simulation is simplified in that it is not a full simulation of the MUSE apparatus but assumes the nominal beam momentum, scattering angle, momentum-detection threshold, and photon calorimeter geometrical acceptance and energy reconstruction. Processes accompanying the passage of incident and outgoing particles through the upstream detector and target materials have not yet fully been considered. Initial tests with a more comprehensive simulation show that this simple approach encompasses all relevant effects and the two approaches give quantitatively similar results.
Uncertainties in the beam momentum and scattering angle are so small that they do not significantly contribute to the uncertainty in the radiative corrections.  The uncertainty in the minimum lepton momentum $p'_\text{min}$, however, contributes strongly to radiative corrections in electron scattering but not in muon scattering.  

\begin{figure}[htbp]
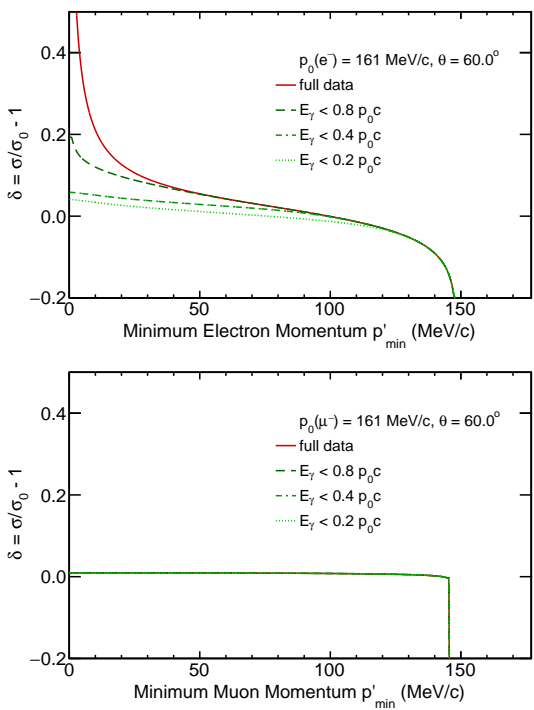

  \centering
  \includegraphics[width=\columnwidth]{{{fig/figures_e-_161.0MeV_060.0deg_Fig.9_lepton_p_delta}}}
  \includegraphics[width=\columnwidth]{{{fig/figures_mu-_161.0MeV_060.0deg_Fig.9_lepton_p_delta}}}
\caption{Results of radiative corrections for MUSE with a beam momentum of 161~MeV/$c$ and a scattering angle of 60$^\circ$.  The top panel shows the results for electron scattering, the bottom panel the results for muons. The curves show the results without (red) and with increasingly strong cuts on initial-state radiation using the calorimeter (green).  Those distributions all overlap in the muon case.
}
\label{fig:delta}
\end{figure}

This result is exemplified in the values for $\delta$ that are shown in Fig.~\ref{fig:delta} as a function of $p'_\text{min}$ for electron (top panel) and muon (bottom panel) beams of a momentum of 161~MeV/$c$ and for a mid-range scattering angle of $60^\circ$. The correction parameter $\delta$ is negative when the scattering cross section is smaller than the Born cross section at high values of $p'_\text{min}$.
The red curves show the results for all scattering events, regardless of the emission of photons.  For $ep$, the full result shows a strong dependence on $p'_\text{min}$ with a steep slope ($\partial \delta / \partial p'_\text{min} >1 \%$ per 1~MeV/$c$) close to the SPS detection threshold of about 14~MeV/$c$. If uncontrolled, the instrumental uncertainty in $p'_\text{min}$ would generate a considerable uncertainty in $\delta$.  

As discussed in Sec.~\ref{sec:tail}, the increase in $\delta$ with a decrease in $p'_\text{min}$ is linked to the increase of the $ep$ cross section with reduced beam momentum after the emission of high-energy initial-state radiation.  The initial-state radiation is strongly forward peaked, and the MUSE calorimeter in the beamline downstream of the target is capable of detecting these bremsstrahlung photons.  The various green curves in Fig.~\ref{fig:delta} show the results for $\delta$ after vetoing an increasing fraction of events with forward going hard photons. The lower the chosen photon-energy cut to suppress the initial-state radiation, the smaller are the radiative corrections and their dependence on $p'_\text{min}$. 

However, as seen in Fig.~\ref{fig:cal}, a sizable fraction of the initial-state-radiation momentum distribution is at low photon momenta.  Selecting the photon-energy cut in that region increases the uncertainty in the experimental cross section and in the corresponding radiative corrections due to uncertainties in the reconstructed photon momentum. To determine the optimum event selection with the smallest overall uncertainty for the radiative corrections for $ep$, we repeated the simulations, systematically increasing the upper limit of the energies of the accepted hard-photons in the forward direction from 0~\% to 100~\% of the beam momentum. 

\begin{figure}[ht]
  \centering
  \includegraphics[width=\columnwidth]{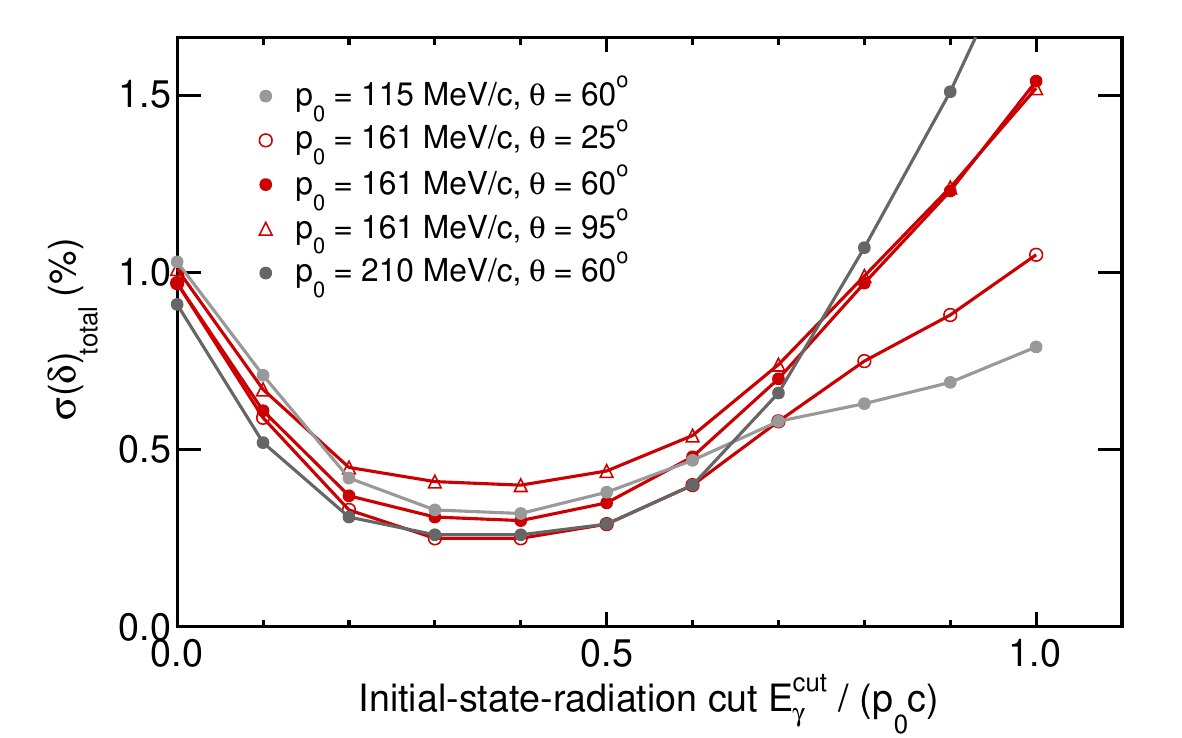}
\caption{Total uncertainties of radiative corrections $\sigma_\delta$ for $ep$ scattering and various beam momenta and scattering angles in MUSE as a function of the photon-energy cut $E_\gamma^\text{cut}$.}
\label{fig:egamma_cut}
\end{figure}

The results are shown in Fig.~\ref{fig:egamma_cut}. We found a shallow minimum of the total uncertainty with a photon energy cut of $E_\gamma < 0.4 p_0c$. Radiative corrections for muons are much smaller than for electrons and nearly independent on $p'_\text{min}$, as shown in the bottom panel of Fig.~\ref{fig:delta}. Because there are not many photons emitted in the forward direction in the $\mu p$ scattering process, the calorimeter cut does not significantly affect the data, and the results of all calculation variations overlap.
\begin{table*}[htb!]
    \caption{Radiative corrections $\delta$ for MUSE.  Values are given for $ep$ and $\mu p$ scattering at various kinematic settings and for fully integrated final-state bremsstrahlung photons (full data) and after suppression of photons in the beam direction with the calorimeter ($E_\gamma < 0.4 p_0c$). The statistical uncertainties in the corrections are of the order $10^{-3}$.}
    \label{tab:rc}
    \begin{center}
    \begin{tabular}{lrrrrrrrrr}
    \hline
    \hline
    $p_0$ (MeV/$c$)               &  115 &  115 &  115 &  161 &  161 &   161 &  210 &  210 &   210 \\
    $\theta$                      &   25$^\circ$ &   60$^\circ$ &    95$^\circ$ &   25$^\circ$ &   60$^\circ$ &    95$^\circ$ &   25$^\circ$ &   60$^\circ$ &    95$^\circ$ \\
    \hline
    \multicolumn{10}{c}{$ep$} \\
    \hline
    $p'_{e,\text{min}}$ (MeV/$c$)       &   14.8 &   13.0 &    14.4 &   14.8 &   13.0 &    14.4 &   14.8 &   13.0 &    14.4 \\
    $\delta_e$  (full data)             &  0.091 &  0.119 &   0.119 &  0.130 &  0.173 &   0.172 &  0.173 &  0.239 &   0.235 \\
    $\delta_e$  ($E_\gamma < 0.4 p_0c$) &  0.026 &  0.042 &   0.049 &  0.028 &  0.049 &   0.060 &  0.030 &  0.056 &   0.070 \\
    \hline
    \multicolumn{10}{c}{$\mu p$} \\
    \hline
    $p'_{\mu,\text{min}}$ (MeV/$c$)       &   84.2 &   82.4 &    85.8 &   84.2 &   82.4 &    85.8 &   84.2 &   82.4 &    85.8 \\
    $\delta_\mu$  (full data)             &  0.001 &  0.005 &   0.003 &  0.004 &  0.009 &   0.008 &  0.006 &  0.010 &   0.010 \\
    $\delta_\mu$  ($E_\gamma < 0.4 p_0c$) &  0.001 &  0.005 &   0.003 &  0.004 &  0.009 &   0.008 &  0.006 &  0.010 &   0.010 \\
    \hline
    \hline
    \end{tabular}
    \end{center}
\end{table*}

Table~\ref{tab:rc} summarises the radiative-correction values at the three MUSE beam momenta and for three scattering angles. The table also indicates the minimum lepton momentum that was used in the determination of the corrections for each scattering angle.  The statistical uncertainties in the corrections are of the order $10^{-3}$.  Suppressing hard initial-state radiation using the photon calorimeter reduces the $ep$ radiative corrections to the Born cross section by a factor of 2.5 to 6, depending on the kinematic setting, to values below 0.1.  The corrections for $\mu p$ scattering are at most 0.01 and independent of the calorimeter response.

The impact of the instrumental uncertainties in the key input parameters $x_i$ of the calculation---the beam momentum, scattering angle, minimum final-state lepton momentum, and the photon-energy cut for forward going photons---was studied by varying that input to determine $\partial \delta / \partial x_i$ and propagate the experimental uncertainties into the uncertainty of the radiative corrections.  The preliminary results for a calorimeter threshold of $E_\gamma < 0.4 p_0 c$ are given in Table~\ref{tab:uncertainties} for $ep$ scattering. 
\begin{table*}[h!]
    \caption{Uncertainties of radiative corrections $\sigma_\delta$ for $ep$ scattering in MUSE, including the various contributions from the experimental uncertainties in the model input parameters $p_0$, $\theta_e$, $p'_\text{min}$, and $E_\gamma$. The values assume a cut on hard photons with $E_\gamma > 0.4 p_0c$ in the MUSE calorimeter. The total uncertainty does not include model uncertainties in ESEPP.}
    \label{tab:uncertainties}
    \begin{center}
    \begin{tabular}{lrrrrrrrrr}
    \hline
    \hline
    $p_0$ (MeV/$c$)               &  115 &  115 &  115 &  161 &  161 &   161 &  210 &  210 &   210 \\
    $\theta$                      &   25$^\circ$ &   60$^\circ$ &    95$^\circ$ &   25$^\circ$ &   60$^\circ$ &    95$^\circ$ &   25$^\circ$ &   60$^\circ$ &    95$^\circ$ \\
%
    \hline
     & \multicolumn{9}{c}{$\sigma_\delta$ (\%)}  \\
    \hline
    $|(\partial \delta_e/\partial p_0)\sigma_{p_0}|$ &
        0.01& 0.01& 0.00& 0.01& 0.00& 0.00& 0.00& 0.03& 0.01 \\
    $|(\partial \delta_e/\partial \theta_e)\sigma_{\theta_e}|$ &
        0.00& 0.00& 0.00& 0.00& 0.00& 0.00& 0.00& 0.00& 0.00 \\
    $|(\partial \delta_e/\partial p'_\text{min})\sigma_{p'_\text{min}}|$ &
        0.05& 0.18& 0.30& 0.03& 0.16& 0.31& 0.02& 0.13& 0.31 \\
    $|(\partial \delta_e/\partial E_\gamma)\sigma_{E_\gamma}|$ &
        0.32& 0.33& 0.33& 0.25& 0.26& 0.26& 0.20& 0.22& 0.22 \\
    \hline
    $\sigma_{\delta_e}$ &
        0.32& 0.38& 0.45& 0.25& 0.30& 0.40& 0.20& 0.26& 0.38 \\
    \hline
    \hline
    \end{tabular}
    \end{center}
\end{table*}
Experimental uncertainties in the beam momentum and scattering angle are too small to affect the radiative corrections significantly.  Contributions to the uncertainty budget from the photon-energy cut in the calorimeter are independent of the scattering angle and decrease from about 0.33 \% to 0.22 \% with increasing beam momentum.  Those uncertainties do contribute to the absolute cross section normalization uncertainty but do not contribute to the uncertainty of the extraction of the proton charge radius which relies on the analysis of the slope of the cross section.  Uncertainties in the knowledge of the electron detection threshold, and thus the lower bound in the integration of Eq.~(\ref{eq:int}), affect the cross section angular distribution.  The contribution to the uncertainty in the radioactive correction is small at forward angles and about 0.3~\% at backward angles.  For $\mu p$ scattering, there are no significant contributions to the uncertainty of radiative corrections from experimental input parameters to the model calculations. The estimated uncertainties of the radiative corrections for $e^+$ ($\mu^+$) scattering is similar to $e^-$ ($\mu^-$) within the statistical precision of our simulations. The results in Table~\ref{tab:uncertainties} are based on the Kelly parametrization of the proton form factors. Simulations with the Dipole form factor lead to identical results in the estimated uncertainties. Other model uncertainties for $ep$ and $\mu p$ scattering, like higher-order corrections to the cross sections, are estimated to have minimal effect on the assessment of the \textit{instrumental} uncertainties in the radiative corrections and are not included in this study. Work is underway to improve the detector calibrations and simulations for the SPS and calorimeter detectors to help reducing instrumental uncertainties in $p'_\text{min}$ and $E_\gamma$, and to minimize MUSE systematic uncertainties. 

\section{Summary}
MUSE is a high-precision experiment to measure the proton charge radius, study possible two-photon exchange mechanisms, and have a direct $\mu / e$ comparison of the elastic cross sections. Without a magnetic spectrometer, MUSE does include a wide range of the final-state lepton momenta in the experimental yield. A dedicated downstream photon detector helps to suppress initial-state radiation effects, and, in turn, reduce instrumental uncertainties to the radiative corrections. Simplified Monte Carlo simulations with the ESEPP event generator show that the radiative corrections $\delta$ to the Born cross section are below 0.1 for $ep$ and 0.01 for $\mu p$ scattering. The total uncertainties of the radiative corrections from the uncertainties in the experimental inputs for electron scattering are smaller than 0.5~\%, including angular dependent contributions that are related to the proton radius extraction, of up to about 0.3~\%. The total uncertainties of the radiative corrections from the uncertainties in the experimental input for muons are negligible. 

\section*{Acknowledgments}
The work was supported by the U.S. National Science Foundation grants PHY-2111050, PHY-2012144, PHY-1812402, PHY-2110229, PHY-2012940, PHY-1505934, PHY-1812402, PHY-2113436, PHY-2209348, PHY-1913653 and HDR-1649909; the U.S. Department of Energy grants DE-SC0012485, DE-SC0016577, DE-AC02-06CH11357, and DE-SC0012589; the Paul Scherrer Institute; and the US-Israel Binational Science Foundation. It was also supported by Sigma Xi grants G2017100190747806 and G2019100190747806.

\bibliography{main}

\end{document}